%% file: main.tex
\documentclass[runningheads]{llncs}
\usepackage[T1]{fontenc}
\usepackage{bbm}
\usepackage{dsfont}
\usepackage{multirow}
\usepackage{graphicx}
\usepackage{amsmath}
\usepackage{algorithm}
\usepackage{algorithmic}
\usepackage{booktabs}
\usepackage{amsfonts}
\usepackage{makecell}
\usepackage{marvosym}

\usepackage{orcidlink}
\renewcommand{\orcidID}[1]{\orcidlink{#1}}
\hypersetup{hidelinks}

\begin{document}
\title{C\textsuperscript{3}: Capturing Consensus with Contrastive Learning in Group Recommendation}
\titlerunning{Capturing Consensus with Contrastive Learning in Group Recommendation}
%
\author{Soyoung Kim\inst{1}\orcidID{0009-0004-4703-2162} \and
Dongjun Lee\inst{2}\orcidID{0009-0004-1397-0855} \and
Jaekwang Kim\inst{3(\textrm{\Letter})}\orcidID{0000-0001-5174-0074}
}
\institute{Department of Applied Artificial Intelligence/Convergence Program for Social Innovation,  Sungkyunkwan University, Seoul, South Korea \\ \email{syfg6716@g.skku.edu}\\ \and
Department of Electrical and Computer Engineering, Sungkyunkwan University, Suwon, South Korea \\ \email{akm5825@g.skku.edu}\\\and
Department of Immersive Media Engineering/Convergence Program for Social Innovation,  Sungkyunkwan University, Seoul, South Korea \\ \email{linux@skku.edu}\\
}
\maketitle              

\begin{abstract}
\input{texs/1-abstract.tex}
\end{abstract}

\input{texs/2-introduction}
\input{texs/3-related-work}
\input{texs/4-methodology}

\input{texs/5-experiments}

\input{texs/6-conclusion}

\input{texs/7-ack}

\bibliographystyle{splncs04}
\bibliography{references}
%




\end{document}

%% file: texs/1-abstract.tex
\label{sec:abstract}
Group recommendation aims to recommend tailored items to groups of users, where the key challenge is modeling a consensus that reflects member preferences. Although several existing deep learning models have achieved performance improvements, they still fail to capture consensus in various aspects: (1) Capturing consensus in small-group ($2\sim5$ members) recommendation systems, which align more closely with real-world scenarios, remains a significant challenge; (2) Most existing models significantly enhance the overall group performance but struggle with balancing individual and group performance. To address these issues, we propose \textbf{C}apturing \textbf{C}onsensus with \textbf{C}ontrastive Learning in Group Recommendation~(C\textsuperscript{3}), which focuses on exploring the consensus behind group decision-making. A Transformer encoder is used to learn both group and user representations, and contrastive learning mitigates overfitting for users with many interactions, yielding more robust group representations. Experiments on four public datasets demonstrate that C\textsuperscript{3} significantly outperforms state-of-the-art baselines in both user and group recommendation tasks.

%% file: texs/2-introduction.tex
\section{Introduction}
\label{sec:introduction}

Group activities, such as traveling with friends or dining with family, are essential in human social life, and these activities are planned through online platforms, such as Mafengwo\footnote{https://www.mafengwo.cn} and Meetup\footnote{https://www.meetup.com}.

Traditional recommendation systems primarily focus on recommending items relevant to individual users. However, such systems are less effective in group recommendation scenarios, where group members have diverse preferences and varied interactions with items. Group recommendation involves challenges such as capturing the noise arising from varying user preferences and interactions~\cite{zhao2024dhmae}.
Prior work has primarily focused on aggregating member preferences to infer group preference~\cite{amer2009group,boratto2011state,cao2018agree,cao2019social,chen2022cuberec}. Early approaches use simple rules such as average~\cite{boratto2011state}, least misery~\cite{amer2009group}, or maximum satisfaction~\cite{baltrunas2010group}.
These heuristic-based methods cannot adapt to the dynamic and diverse opinions of group members~\cite{wu2023consrec}. With the advancement of deep learning, attention-based models have been introduced. For example, AGREE{\cite{cao2018agree}} utilizes attention mechanisms to dynamically adjust the influence of group members. However, assigning higher weights to users with frequent interactions lead to overfitting, especially in small-group, where recommendations become overly focus on highly interactive individuals. Recently, hypergraph neural network-based (HGNN-based) methods, which allow a single node to connect to multiple edges, have been applied to group recommendation systems. This approach is particularly suitable for scenarios involving groups composed of multiple users and has demonstrated significant performance improvements in this domain~\cite{chen2022cuberec,wu2023consrec}.

\begin{figure}[!t]
    \centering
    \includegraphics[width=0.7\linewidth]{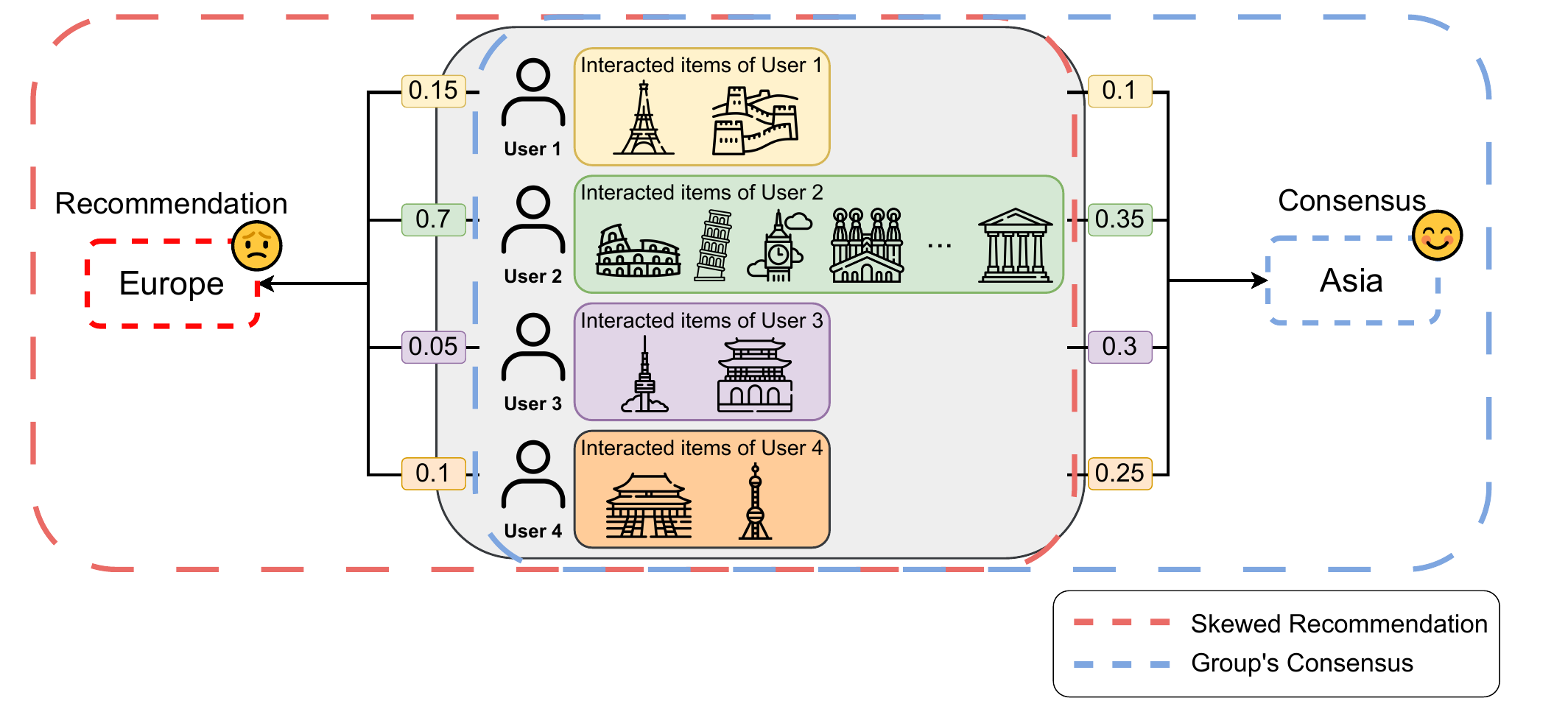}
    \caption{Comparison of group recommendation scenarios
    a red-dotted example where one dominant user shifts the choice to Europe, and a blue-dotted example where full preference aggregation yields an Asia-leaning consensus.
    }
    \label{fig:consensus}
\end{figure}

Despite these advancements, challenges remain. \textbf{(1) The challenge of capturing small-group~(i.e., between 2 and 5 members)  consensus.} Consensus reflects a group's final agreement, integrating the diverse preferences of its members and guiding decision making~\cite{wu2023consrec}. 
In this study, we regard a model as effectively capturing consensus if the group representation for the target item remains stable even when some users are randomly masked. 
As illustrated in Fig.~\ref{fig:consensus}, consider a small group that has rated visited places. Because most members frequently travel to destinations in Asia, the group’s consensus preference for an overseas trip should naturally lean toward Asia. However, attention-based methods may overfit to a highly interactive member and skew recommendations toward Europe.
Recent HGNN-based models have been extensively studied to capture group consensus~\cite{wu2023consrec,xu2024aligngroup}.
ConsRec~\cite{wu2023consrec} was designed to comprehensively capture group consensus by using hypergraph convolutional operations, which enable the generation of rich member representations.
However, most HGNN-based models exhibit performance degradation on datasets with high interaction but small-group~\cite{chen2022cuberec,wu2023consrec}. This is because they fail to provide diverse recommendations that reflect the preferences of all group members. 

\textbf{(2) Contrastive learning without leveraging the group’s original preference representation.}
Group-item interactions are often sparse, posing substantial challenges in effectively learning group preferences. To address this, contrastive learning has been adopted in group recommendation for data sparsity problem~\cite{zhao2024dhmae}. For instance, CubeRec~\cite{chen2022cuberec} utilizes hypercube modeling integrated with contrastive self-supervision to address data sparsity issues. The Transformer encoder-based model GBERT~\cite{zhang2022gbert} employs contrastive learning by generating fake groups to compare with real groups. However, the effectiveness of contrastive learning relies on high-quality data augmentation and the accurate generation of contrasting representations, making it challenging to generalize these methods to diverse group recommendation scenarios~\cite{zhao2024dhmae}.
Furthermore, these methods fail to effectively utilize the representation of the original group, which can result in overfitting to the remaining users when the group size temporarily changes.

\textbf{(3) Difficulty of generalization.} 
Most group recommendation systems focus on improving group performance, even if it involves reducing recommendation performance for individual users~\cite{xu2024aligngroup}. Since group recommendation aggregates the preferences of individual users, an effective group recommendation system must not only perform well for groups but also offer accurate recommendations for individual users. It can be argued that such a model is effective and demonstrates generalization ability.

To address these challenges, we present \textbf{C}apturing \textbf{C}onsensus with \textbf{C}ontrastive Learning in Group Recommendation, namely C\textsuperscript{3}. 
C\textsuperscript{3} employs a Transformer encoder to model both user–item and group–item interactions, enabling a generalized recommendation system that mitigates performance degradation in user tasks.
In particular, we design a recommendation loss with three components: positive, negative, and margin, as well as a contrastive loss based on an augmentation strategy.
Positive loss emphasizes items interacted with by a user or group, with logarithmic scaling to regulate its magnitude. Negative loss suppresses predictions for non-interacted items via exponential scaling, while margin loss explicitly widens the positive–negative gap to enforce stronger separation.
We also promote group consensus by creating two augmented views of each group through random user masking. 
In each batch, the two corrupted views form a positive pair against all others, pulling positives together and pushing negatives apart. This stabilizes group representations under member dropouts, prevents overfitting to highly active users, and better captures whole-group preferences.

%% file: texs/3-related-work.tex
\begin{figure*}[!t]
    \centering
    \includegraphics[width=0.8\linewidth]{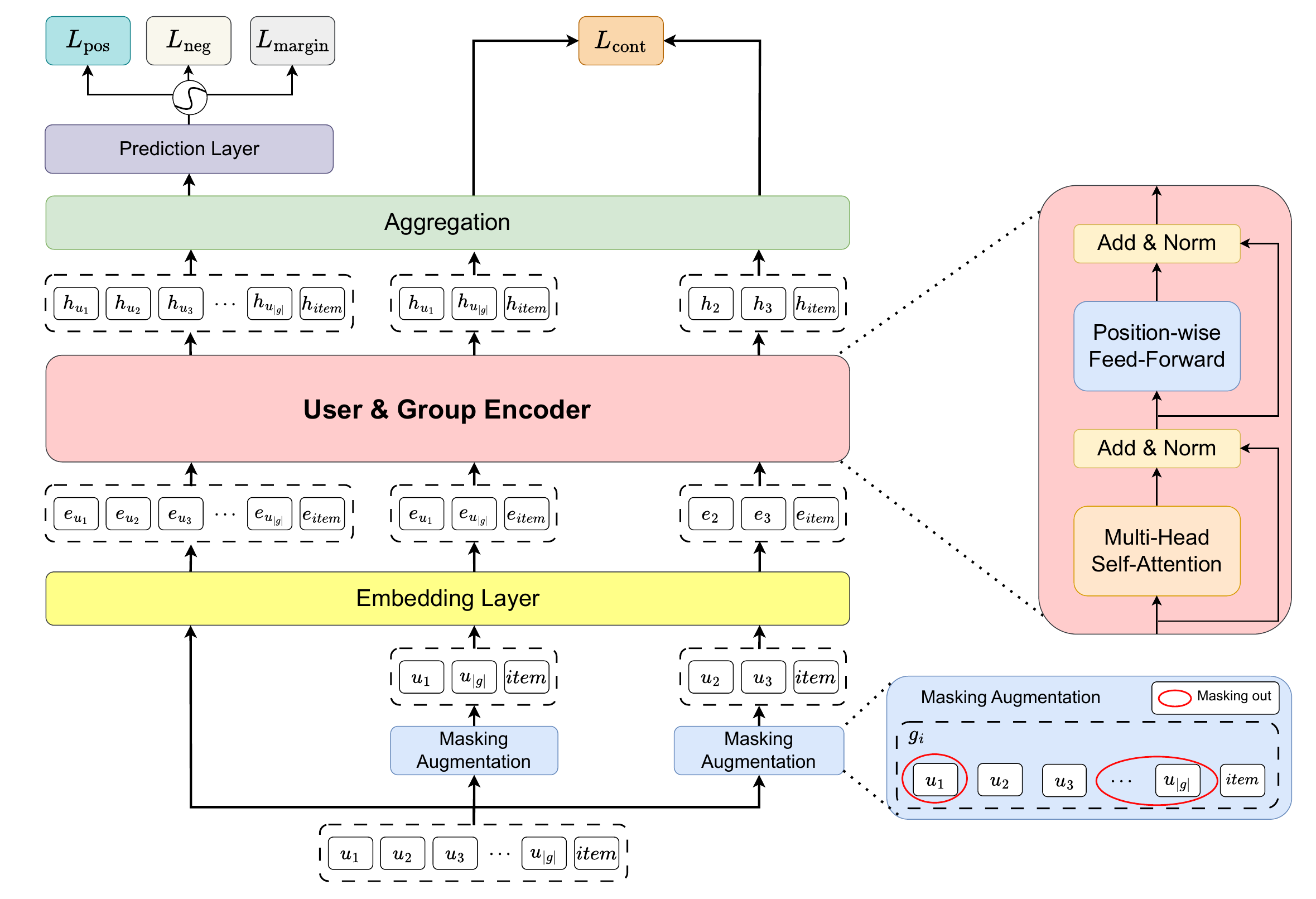}
    \caption{
    Overview of our proposed C\textsuperscript{3}, consisting of the transformer encoder, margin loss, and contrastive loss.
}
    \label{fig:overview}
\end{figure*}

%% file: texs/4-methodology.tex
\section{Method}
\label{sec:METHODOLOGY} 
\subsection{Task Definition}
\label{sec:Task Definition}

Let $\mathcal{U}$, $\mathcal{I}$, and $\mathcal{G}$ denote the user, item, and group sets: $\mathcal{U}=\{u_1, u_2, \dots,u_M\}$, $\mathcal{I}=\{i_1, i_2, \dots,i_N\}$, and $\mathcal{G}=\{g_1, g_2, \dots,g_K\}$, where $M$, $N$, and $K$ are their sizes. We consider two interaction types: user–item and group–item. For group $g_k$, its interacted items are $\mathcal{Y}_k=\{i_1, i_2, \dots,i{|\mathcal{Y}_k|}\}$ with $i_n\in\mathcal{I}$. Given a target group $g_t$, we predict a ranked list of items the group is likely to prefer~\cite{wu2023consrec}.

\subsection{C\textsuperscript{3} Framework}
\label{sec:Framework}
\subsubsection{Embedding Layer.}
\label{sec:embedding}
We embed each user and item into $d$-dimensional vectors using $\mathbf{E}_u\in\mathbb{R}^{M\times d}$ and $\mathbf{E}_i\in\mathbb{R}^{N\times d}$. For a group $g=\{u_1^g, u_2^g, \dots,u_{|g|}^g\}$, each member $u_m^g$ is mapped to $\mathbf{e}_m^g$ via $\mathbf{E}_u\cdot u_m^g$, and each item to $\mathbf{e}_{\text{item}}$. The input to C\textsuperscript{3} is the concatenation of user and item embeddings, formulated as:
\begin{equation}
\label{eq1}
    \textbf{H}^0_{g,i} = \text{Concat}(\{\mathbf{e}_{u_j^g}\}_{j=1}^{|g|}, e_\text{item}),
\end{equation}
where $\textbf{H}^0_g$ represents the initial hidden state. In the case of user recommendation, where $|g| = 1$, the group consists of a single user.
\subsubsection{User \& Group Encoder.}
The encoder in Fig.~\ref{fig:overview} is built from stacked Transformer encoder layers~\cite{devlin2018bert}. Each layer contains two submodules: a multi-head self-attention (MHSA) block and a position-wise feed-forward (FF) network. MHSA uses self-attention to capture dependencies among tokens in a sequence, and its effectiveness has been widely validated in prior work. In our model, MHSA is used to model interactions between group members and candidate item, enabling the encoder to learn representations that reflect group preference dynamics:
\begin{equation}
\begin{aligned}
    \text{MHSA}(\textbf{H}^l) = \text{Concat}(\text{head}_1, \dots, \text{head}_h)\textbf{W}^O, \\
    \text{where,}\,\,\text{head}_i = \text{Attention}(\textbf{H}^l \textbf{W}_i^Q, \textbf{H}^l \textbf{W}_i^K, \textbf{H}^l \textbf{W}_i^V), \\
    \text{Attention}(\textbf{Q}, \textbf{K}, \textbf{V}) = \text{softmax}\left(\frac{\textbf{QK}^\top}{\sqrt{d_k}}\right)\textbf{V},
\end{aligned}
\end{equation}
Here, $\textbf{W}_i^Q$, $\textbf{W}_i^K$, $\textbf{W}_i^V \in \mathbb{R}^{d \times \frac{d}{h}}$, $\textbf{W}^O \in \mathbb{R}^{d \times d}$ are trainable parameters.
Residual connections and layer normalization~\cite{ba2016layer} are applied as in standard Transformers.
\begin{equation}
\begin{aligned}
    \mathbf{\tilde{H}}^l = \text{LN}(\mathbf{H}^l + \text{MHSA}(\mathbf{H}^l)),\\
    \mathbf{H}^{l+1} = \text{LN}(\mathbf{\tilde{H}}^l + \text{FF}(\mathbf{\tilde{H}}^l)), \\
\end{aligned}
\end{equation}
This architecture assigns high weights based on members' contribution to the final prediction. 
\subsubsection{Prediction Layer.}
The prediction layer estimates the probability that a user or group will prefer an item. The encoder outputs hidden vectors for the members and the item, which are merged into a single representation via mean pooling:
\begin{equation}
\label{eq6}
    \mathbf{h}_{g,i} = \frac{1}{|g|+1} \left(\sum_{j=1}^{|g|} \textbf{h}_{u_j} + \textbf{h}_{item}\right),
\end{equation}
where $\mathbf{h}_{g,i}$ denotes the aggregated group representation. We then apply a linear layer followed by a sigmoid to obtain the interaction score:
\begin{equation}
\label{eq7}
    s_{g,i} = \sigma(\mathbf{W} \mathbf{h}_{g,i} + \textbf{b})
\end{equation}
\subsection{Optimization for Recommendation Task}
\label{sec:Optimization}
We treat items interacted with by a user or group as positive, and all others as negative. For training, we use a recommendation loss composed of positive, negative, and margin terms, defined as follows. The positive loss encourages high probability for positive items:
\begin{equation}
\mathcal{L}_\text{pos} = \frac{1}{|\mathcal{P}|} \sum_{g \in \mathcal{G}, i \in \mathcal{P}} -\log(s_{g,i} + \epsilon),
\end{equation}
where $\mathcal{P}$ is the set of positive items,
and $\epsilon$ is a negligible constant for numerical stability. 
Logarithmic scaling is applied to this loss function to ensure that lower scores assigned to positive items result in higher penalties. 
Negative loss penalizes predictions for negative items, thus ensuring that negative items receive lower probability in the predictions.
\begin{equation}
\mathcal{L}_\text{neg} = \frac{1}{|\mathcal{N}|} \sum_{g \in \mathcal{G}, j \in \mathcal{N}} \left[\exp(s_{g,j}) - 1\right],
\end{equation}
$\mathcal{N}$ is the set of negative samples. Additionally, margin loss enforces a difference between the predictions of positive and negative samples, ensuring that positive items rank higher than negative items in the predictions:
\begin{equation}
\mathcal{L}_\text{margin} = \frac{1}{|\mathcal{P}|} \sum_{g \in \mathcal{G}, p \in \mathcal{P}} \frac{1}{|\mathcal{N}|} \sum_{n \in \mathcal{N}} \max\left(0, \delta - (s_{g,p} - s_{g,n})\right),
\end{equation}
$\delta$ is the margin hyperparameter. 
Unlike Bayesian Personalized Ranking (BPR)~\cite{rendle2012bpr}, which only adjusts pairwise item ordering, proposed losses adjust predicted probabilities to clearly distinguish positive from negative items, thereby reflecting absolute scores. Moreover, the margin term encourages the model to maintain a relative gap between positive and negative probabilities. 
Finally, the total loss is computed as follows:
\begin{equation}
\mathcal{L}_\text{main} = \alpha \left(\mathcal{L}_\text{pos} + \mathcal{L}_\text{neg}\right) + (1 - \alpha) \mathcal{L}_\text{margin}
\end{equation}
where $\alpha$ is a hyperparameter that mediates the balance between two losses.

\subsection{Contrastive Learning}
\label{sec:Contrastive learning}

\subsubsection{Augmentation.}
\label{sec:Augmentation}

A group consists of multiple users whose preferences are aggregated with a encoder. However, since self-attention favors users with stronger interactions, the model may overweight highly active members and produce biased recommendations.
To address this, we design a masking augmentation that preserves  members’ preferences and produces robust representations even with missing users. In training, we randomly mask users and generate two independent augmented views per group. The two views from the same original group constitute a positive pair, while views from different groups are treated as negatives. This augmentation is formulated as:

\begin{equation} 
\label{eq12}
    g' = g \setminus \mathcal{M}, \quad \mathcal{M} \subseteq g, \quad |\mathcal{M}| = \lfloor r \cdot |g| \rfloor
\end{equation}
$g'$ represents the augmented group.
$\mathcal{M}$ denotes the selected subset of $g$ for masking, and $r$ indicates the masking ratio. To distinguish between augmented samples, 
for the i-th group $g_i$, the corresponding augmented samples are denoted as $g'_{2j}$ and $g'_{2j+1}$. In this module, we randomly select a subset of $\mathcal{M}$ for masking. This randomness prevents the learned representations from overfitting to specific users and helps avoid trivial collapse in contrastive learning. Additionally, we introduce a threshold hyperparameter to regulate the augmentation. 
Augmentation is applied only to groups above a size threshold to avoid excessive distortion in very small groups.

\subsubsection{Self-supervised Learning Signal.}
\label{sec:self-supervised learning signal}
As discussed in the previous section, augmented samples pass through all layers to obtain the hidden representations $\textbf{h}_{g'_{2j}}$ and $\textbf{h}_{g'_{2j+1}}$. Then, we compute the contrastive loss using the InfoNCE~\cite{oord2018representation}:
\begin{equation}
\label{eq13}
    \mathcal{L}_{\textbf{Cont}}(g'_{2j}, g'_{2j+1}) = -\text{log}\frac{\text{exp}(\text{sim}(\textbf{h}_{g'_{2j}}, \textbf{h}_{g'_{2j+1}})/\tau)}{\sum_{k=1}^{2B} \mathbbm{1}_{[j \neq k]}\text{exp}(\text{sim}(\textbf{h}_{g'_{2j}}, \textbf{h}_{g'_k})/\tau) }
\end{equation}
where $B$ is the batch size and $\tau$ is the temperature. 
This loss pulls positive pairs together and pushes negatives apart within each mini-batch, enabling the encoder to learn similar representations for augmented groups with removed users and thus maintain stable group representations even when highly engaged members are masked.
Group-preferred items should reflect the consensus of all members. Therefore, even if excluding some users changes the top-ranked item as group composition varies, the model should still assign high scores to previously preferred items.
This loss is incorporated into main loss with a scaling factor $\beta$: 

\begin{equation}
    \mathcal{L} = \mathcal{L}_\text{main} + \beta \cdot \mathcal{L}_\text{cont}
\end{equation}

%% file: texs/5-experiments.tex
\section{Experiments}
\label{sec:Experiments}

\begin{table}[!t]
\centering
\caption{\textbf{The statistics of the datasets.}}
\label{tab:dataset_statistics}
\centering

\scalebox{0.8}{ 
\begin{tabular}{l|rrrrrr}
\toprule
Dataset & \#Users & \#Items & \#Groups & \makecell{\#U-I interactions} & \makecell{\#G-I interactions} & \makecell{Avg. group size}\\

\midrule
\textbf{Mafengwo}
& 5,275
& 1,513
& 995
& 39,761
& 3,595
& 7.19
\\
\textbf{CAMRa2011}
& 602
& 7,710
& 290
& 116,344
& 145,068
& 2.08
\\
\textbf{MovieLens}
& 943
& 1,682
& 5,000
& 99,057
& 88,705
& 4.27
\\
\textbf{MafengwoS}
& 11,027
& 1,236
& 1,215
& 8,237
& 2,468
& 12.34
\\
\bottomrule
\end{tabular}}
\end{table}


\subsection{Experimental Setup} 
\subsubsection{Datasets.}
Mafengwo is a tourism platform where users record visited locations and form travel groups. CAMRa2011~\cite{cao2018agree} includes 
household movie ratings, containing group preferences.
MovieLens, designed in previous group recommendation research~\cite{zhao2024dhmae}, is constructed by randomly selecting users as members of a group.
MafengwoS~\cite{wu2023consrec} is a reconstructed Mafengwo dataset that includes semantic details such as item names.
Table~\ref{tab:dataset_statistics}
summarizes the dataset statistics.

\subsubsection{Evaluation Metrics.}  
Following prior settings~\cite{cao2018agree,cao2019social}, 
We evaluate the top-K recommendation performance using hit ratio at K~(HR@K) and normalized discounted cumulative gain at K~(NDCG@K)~\cite{jarvelin2002cumulated}. We reports results for HR@1/5/10 and NDCG@5/10.
Best performance is in bold and second-best is underlined.

\subsubsection{Baselines.}

\begin{table*}[!t]
\centering
\caption{Performance comparison over all baselines on group recommendation task.}
\label{tab:performance_group}
\resizebox{1.0\linewidth}{!}{
\begin{tabular}{c|l|ccccccc|c|r} 
\toprule

Datasets
& Metrics
& NCF
& AGREE
& GroupIM
& S\textsuperscript{2}-HHGR
& CubeRec
& ConsRec
& AlignGroup
& C\textsuperscript{3}
& Improv.
\\
\hline
\multirow{5}{*}{\textbf{Mafengwo}}
& HR@1
& 0.3242
& 0.4563
& 0.3387
& 0.6809
& 0.6312
& 0.6231
& \underline{0.6744}
& \textbf{0.8613}
& 27.71\%
\\
& HR@5
& 0.4291
& 0.7045
& 0.6412
& 0.7568
& 0.8533
& 0.8824
& \textbf{0.9548}
& \underline{0.8864}
& -
\\
& HR@10
& 0.6181
& 0.7769
& 0.7367
& 0.7779
& 0.9025
& 0.9056
& \textbf{0.9628}
& \underline{0.9115}
& -
\\
& NDCG@5
& 0.3405
& 0.5892
& 0.4986
& 0.7322
& 0.7572
& 0.7697
& \underline{0.8343}
& \textbf{0.8711}
& 4.41\%
\\
& NDCG@10
& 0.4020
& 0.6126
& 0.5295
& 0.7391
& 0.7734
& 0.7794
& \underline{0.8369}
& \textbf{0.8796}
& 5.10\%
\\
\hline
\multirow{5}{*}{\textbf{CAMRa2011}}
& HR@1
& 0.1957
& 0.3097
& 0.2083
& 0.1600
& 0.2110
& 0.1966
& \underline{0.8048}
& \textbf{0.9255}
& 15.00\%    
\\
& HR@5
& 0.5803
& 0.5910
& 0.6152
& 0.6131
& 0.6435
& 0.6366
& \underline{0.8062}
& \textbf{0.9262}
& 14.88\%
\\
& HR@10
& 0.7693
& 0.7807
& 0.7993
& 0.8152
& 0.8262
& 0.8145
& \underline{0.8324}
& \textbf{0.9269}
& 11.35\%
\\
& NDCG@5
& 0.3896
& 0.4016
& 0.4164
& 0.3882
& 0.4334
& 0.4218
& \underline{0.8055}
& \textbf{0.9258}
& 14.93\%
\\
& NDCG@10
& 0.4448
& 0.4530
& 0.4763
& 0.4547
& 0.4932
& 0.4793
& \underline{0.8135}
& \textbf{0.9260}
& 13.83\%
\\
\hline
\multirow{5}{*}{\textbf{MovieLens}}
& HR@1
& 0.2230
& 0.4349
& 0.1512
& 0.0280
& 0.3220
& 0.4546
& \underline{0.7624}
& \textbf{0.9258}
& 21.43\%    
\\
& HR@5
& 0.3991
& 0.7816
& 0.6177
& 0.2151
& 0.7282
& 0.7615
& \underline{0.8268}
& \textbf{0.9266}
& 12.07\%
\\
& HR@10
& 0.5886
& 0.8871
& 0.8232
& 0.3622
& 0.8615
& 0.8531
& \underline{0.8914}
& \textbf{0.9318}
& 4.53\%
\\
& NDCG@5
& 0.3899
& 0.6176
& 0.3869
& 0.1203
& 0.5378
& 0.6213
& \underline{0.7941}
& \textbf{0.9261}
& 16.62\%
\\
& NDCG@10
& 0.4561
& 0.6519
& 0.4545
& 0.1675
& 0.5812
& 0.6512
& \underline{0.8148}
& \textbf{0.9278}
& 13.87\%
\\
\hline
\multirow{5}{*}{\textbf{MafengwoS}}
& HR@1
& 0.0366
& 0.1581
& 0.2526
& 0.0516
& \underline{0.4089}
& 0.3436
& 0.3471
& \textbf{0.4227}
& 3.37\%
\\
& HR@5
& 0.0728
& 0.4485
& 0.5464
& 0.1787
& 0.6289
& \underline{0.6443}
& 0.6392
& \textbf{0.7354}
& 14.14\%
\\
& HR@10
& 0.1965
& 0.5997
& 0.6632
& 0.2698
& 0.6993
& 0.7106
& \underline{0.7680}
& \textbf{0.8076}
& 5.16\%
\\
& NDCG@5
& 0.0978
& 0.3062
& 0.4038
& 0.1169
& \underline{0.5052}
& 0.5051
& 0.5012
& \textbf{0.5771}
& 14.23\%
\\
& NDCG@10
& 0.1064
& 0.3555
& 0.4422
& 0.1464
& \underline{0.5482}
& 0.5362
& 0.5436
& \textbf{0.6012}
& 9.67\%
\\
\bottomrule
\end{tabular}
}
\end{table*}

\begin{table*}[!t]
\centering
\caption{Performance comparison of all methods on the user recommendation task.}
\label{tab:performance_user}

\resizebox{1.0\linewidth}{!}{
\begin{tabular}{c|l|ccccccc|c|r} 
\toprule

Datasets
& Metrics
& NCF
& AGREE
& GroupIM
& S\textsuperscript{2}-HHGR
& CubeRec
& ConsRec
& AlignGroup
& C\textsuperscript{3}
& Improv.
\\
\hline
\multirow{5}{*}{\textbf{Mafengwo}}

&HR@1
&0.2386
&0.2816
&0.2406
&0.2497
&0.0393
&0.6009
&\underline{0.6693}
&\textbf{0.9282}
&38.68\%
\\

&HR@5
&0.6363
&0.6790
&0.6237
&0.6761
&0.1887
&0.7720
&\underline{0.8033}
&\textbf{0.9373}
&16.68\%
\\

&HR@10
&0.7417
&0.7834
&0.7674
&0.7976
&0.3478
&0.8381
&\underline{0.8541}
&\textbf{0.9521}
&11.47\%
\\

&NDCG@5
&0.5432
&0.4951
&0.4447
&0.4781
&0.1121
&0.6877
&\underline{0.7377}
&\textbf{0.9321}
&26.35\%
\\

&NDCG@10
&0.5733
&0.5293
&0.4911
&0.5176
&0.1636
&0.7095
&\underline{0.7541}
&\textbf{0.9368}
&24.23\%
\\

\hline
\multirow{5}{*}{\textbf{CAMRa2011}}
& HR@1
&0.3285
&0.3568
&0.2113
&0.1741
&0.1641
&0.2199
&\underline{0.8209}
&\textbf{0.9485}
&15.54\%
\\

&HR@5
&0.6119
&0.6243
&0.6229
&0.6375
&0.6276
&0.6774
&\underline{0.8239}
&\textbf{0.9485}
&15.12\%
\\

&HR@10
&0.7894
&0.7967
&0.7993
&0.8239
&0.8215
&\underline{0.8515}
&0.8348
&\textbf{0.9502}
&11.59\%
\\

&NDCG@5
&0.4018
&0.4217
&0.4234
&0.4105
&0.4054
&0.4531
&\underline{0.8222}
&\textbf{0.9485}
&15.36\%
\\

&NDCG@10
&0.4535
&0.4827
&0.4810
&0.4718
&0.4686
&0.5100
&\underline{0.8256}
&\textbf{0.9490}
&14.95\%
\\

\hline
\multirow{5}{*}{\textbf{MovieLens}}
&HR@1
&0.0258
&0.0817
&0.0477
&0.0668
&0.0371
&0.1644
&\underline{0.4677}
&\textbf{0.6723}
&43.75\%
\\

&HR@5
&0.1278
&0.3128
&0.2238
&0.3044
&0.1474
&0.3139
&\underline{0.4783} 
&\textbf{0.6734}
&40.79\%
\\

&HR@10
&0.2869
&0.4836
&0.3701
&0.4899
&0.2566
&0.4846
&\underline{0.5281}
&\textbf{0.6851}
&29.73\%
\\

&NDCG@5
&0.0775
&0.1955
&0.1329
&0.1825
&0.0918
&0.2354
&\underline{0.4722}
&\textbf{0.6729}
&42.50\%
\\

&NDCG@10
&0.1926
&0.2507
&0.1793
&0.2425
&0.1269
&0.2902
&\underline{0.4878}
&\textbf{0.6766}
&38.70\%
\\

\hline
\multirow{5}{*}{\textbf{MafengwoS}}

&HR@1
&0.0046
&0.1225
&0.1004
&0.0783
&0.0735
&\underline{0.1434}
&0.0741
&\textbf{0.2915}
& 103.28\%
\\

&HR@5
&0.0929
&0.3554
&0.3220
&0.2903
&0.2127
&\underline{0.3811}
&0.2784
&\textbf{0.3967}
& 4.09\% \\ 

&HR@10
&0.1565
&\underline{0.5137}
&0.4851
&0.4361
&0.3345
&0.5054
&0.4265
&\textbf{0.5538}
& 7.81\%
\\ 

&NDCG@5
&0.0688
&0.2419
&0.2124
&0.1843
&0.1430
&\underline{0.2645}
&0.1766
&\textbf{0.3398}
& 28.47\%
\\

&NDCG@10
&0.0894
&0.2926
&0.2649
&0.2318
&0.1826
&\underline{0.3043}
&0.2242
&\textbf{0.3904}
& 28.29\%

\\
\bottomrule
\end{tabular}
}
\end{table*}

We evaluate our model \footnote{The code is available on https://github.com/soso-young/C3.} against state-of-the-art baselines:\\
\textbf{Classical neural network-based model}: (1) \textbf{NCF}~\cite{he2017neural} treats each group as a single virtual user for making predictions. \\
\textbf{Attention-based models}:
(1) \textbf{AGREE}~\cite{cao2018agree} employs an attention mechanism to dynamically aggregate user preferences within a group.
(2) \textbf{GroupIM}~\cite{sankar2020groupim} aggregates user preferences via attention mechanism and applies self-supervised learning to address data sparsity. \\
\textbf{Graph neural network~(GNN)-based models}:
(1) \textbf{S\textsuperscript{2}-HHGR}~\cite{zhang2021double} employs hierarchical hypergraph to capture interactions within and across groups.
(2) \textbf{CubeRec}~\cite{chen2022cuberec} uses hypercubes to adaptively aggregate group members’ interests through geometric expressiveness.
(3) \textbf{ConsRec}~\cite{wu2023consrec} captures group consensus by jointly modeling member, item, and group perspectives.
(4) \textbf{AlignGroup}~\cite{xu2024aligngroup} captures group consensus via hypergraph and leverages self-supervised learning to reduce the gap between group consensus and individual preferences. DHMAE~\cite{zhao2024dhmae} is excluded due to an issue that produced incorrect predictions\footnote{https://github.com/ICharlotteI/DHMAE/issues/3}.






\subsubsection{Implementation Details.} 
We optimize our method with Adam~\cite{kingma2014adam}.
The embedding dimension sets to 32, and the model uses 3 layers with a 0.2 dropout rate. We set $\tau$ = 1.0, margin $\delta$ = 1, and $\alpha$ = 0.5. The threshold is tuned from $\{3,5,7,9\}$, the mask ratio from $\{0.2,0.4,0.6,0.8\}$, and $\beta$ from $\{0.025,0.05,0.075, \\ 0.1\}$. We adopt baseline hyperparameters from the original papers.

\subsection{Overall Performance Comparison}
\label{sec:RQ1}
\textbf{Group recommendation task.}
C\textsuperscript{3} achieves significant improvements in group recommendation performance, particularly in @1 and the NDCG metrics. 
Furthermore, the improvements are particularly evident on small-group datasets such as CAMRa2011 and MovieLens. Our approach learns through masking group members to capture the diverse preferences of all users within the group. It indicates that learning the preferences of all users more uniformly, rather than relying on specific individual, is crucial to precisely modeling group representation. In contrast, GNN-based models exhibit a significant decline in performance across nearly all metrics on small-group datasets.
This suggests that GNN are well-suited for group scenarios; however, their performance degrades when groups are small, as the reduced node–edge connectivity limits their effectiveness.
\\
\textbf{User recommendation task.}
In the user recommendation task,(illustrated in Table~\ref{tab:performance_user}), C\textsuperscript{3} demonstrates outstanding performance across all metrics and datasets. 
The Transformer architecture is key to this result. Unlike a Transformer model~\cite{zhang2022gbert} designed to take only user embeddings as input, our model takes both user and item inputs. This design enables the model, when learning user representations, to capture a user’s individual preferences and that user’s preferences within the group, allowing the two tasks to complement each other. 
It is noted that robust user representations improve group modeling and recommendation performance for both users and groups. Thus, C\textsuperscript{3} can be considered a generalized group recommendation system.

\subsection{Ablation Study}
\label{sec:RQ2}
\begin{table}[!t]
\centering
\caption{Ablation study on the contribution of our framework's loss functions. 
}
\label{tab:ablation}
\resizebox{0.8\linewidth}{!}{
\setlength{\tabcolsep}{10pt} 
\begin{tabular}{c|c|l|ccc} 
\toprule

\textbf{Task}
& \textbf{Dataset}
& \textbf{Metrics}
& \textbf{w/o. Margin}
& \textbf{w/o. Cont}
& \textbf{C\textsuperscript{3}}
\\
\hline
\multirow{4}{*}{\textbf{Group}}
& \multirow{2}{*}{\textbf{CAMRa2011}}
& HR@10
& 0.8903
& 0.9110
& \textbf{0.9269}
\\
&
& NDCG@10
& 0.8846
& 0.9082
& \textbf{0.9260}
\\
\cmidrule{2-6}
& \multirow{2}{*}{\textbf{MovieLens}}
& HR@10
& 0.9022
& 0.9163
& \textbf{0.9318}
\\
&
& NDCG@10
& 0.8804
& 0.9055
& \textbf{0.9278}
\\

\hline
\multirow{4}{*}{\textbf{User}}
& \multirow{2}{*}{\textbf{CAMRa2011}}
& HR@10
& 0.9173
& 0.9286
& \textbf{0.9502}
\\
&
& NDCG@10
& 0.9188
& 0.9249
& \textbf{0.9490}
\\
\cmidrule{2-6}
& \multirow{2}{*}{\textbf{MovieLens}}
& HR@10
& 0.6066
& 0.6617
& \textbf{0.6851}
\\
&
& NDCG@10
& 0.5873
& 0.6546
& \textbf{0.6766}
\\

\bottomrule 
\end{tabular} 
}
\end{table}
\textbf{Without Margin Loss (w/o. Margin).}
Margin loss showed the largest improvement, with 5.38\% in group and 15.21\% in user, respectively. Since margin loss focuses on maximizing the interval between positive and negative items, it directly impacts the ranking list. Specifically, it showed significant improvements in user recommendation.\\
\textbf{Without Contrastive Loss (w/o. Cont).} Our contrastive loss improves performance by 2.46\% for groups and 3.54\% for users. Although designed to capture diverse group preferences, its gain in user recommendation likely stems from the Transformer.
The Transformer inherently assigns attention to interactive members through relationship among user(s) and item, while contrastive loss encourages learning on less interactive members, leading to enhanced user representation. Overall, these losses strengthen the efficiency of C\textsuperscript{3}.

\subsection{Capturing a Consensus}
\label{sec:RQ3}
\begin{figure}[!t] 
    \includegraphics[width=1.0\linewidth]{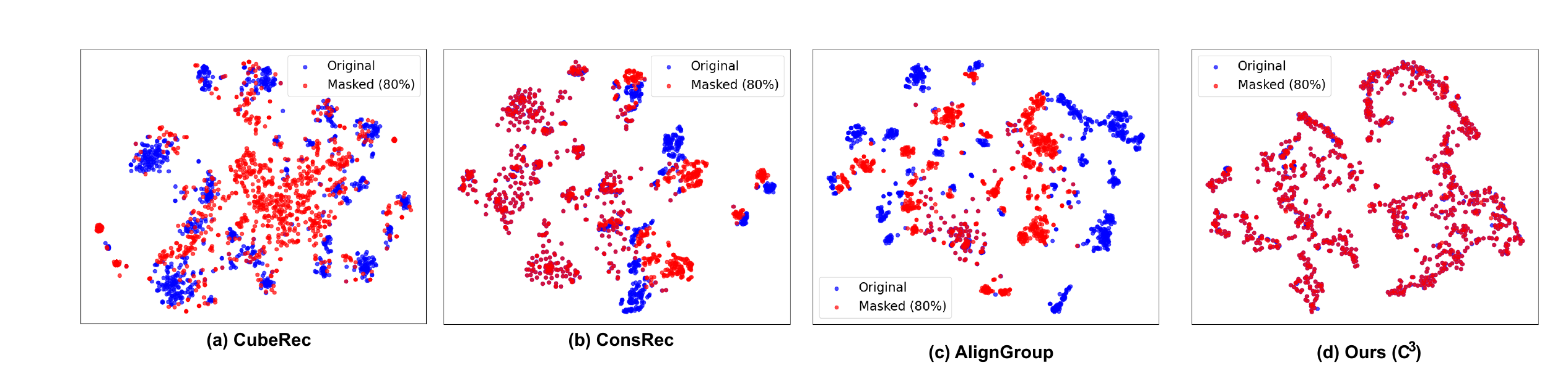}
    \caption{
    Visualization of the original representations (blue dots) and masked representations (red dots), where 80\% of random users in a group are removed on Mafengwo.
    }
    \label{fig:rq4}
\end{figure}

This evaluation demonstrates the extent to which the model preserves the robustness of the group representation, which represents the aggregated preferences of group members for the target item. To illustrate this, we visualize group embeddings using t-SNE~\cite{van2008visualizing} comparing original groups with versions where 80\% of users are removed.
We experiment with models designed to capture group consensus~(ConsRec~\cite{wu2023consrec} and AlignGroup~\cite{xu2024aligngroup}) and a self-supervised learning-based model~(CubeRec~\cite{chen2022cuberec}). As shown in Fig.~\ref{fig:rq4}, ConsRec and AlignGroup show minimal drift under user removal, while CubeRec preserves some structure but with noticeable changes. In contrast, C\textsuperscript{3} produces representations nearly identical to the originals, demonstrating that it reflects consensus in recommendations.

\subsection{Effect of Different Group Sizes}
\label{sec:RQ4}
\begin{figure}[!t] 
    \centering
    \includegraphics[width=0.8\linewidth]{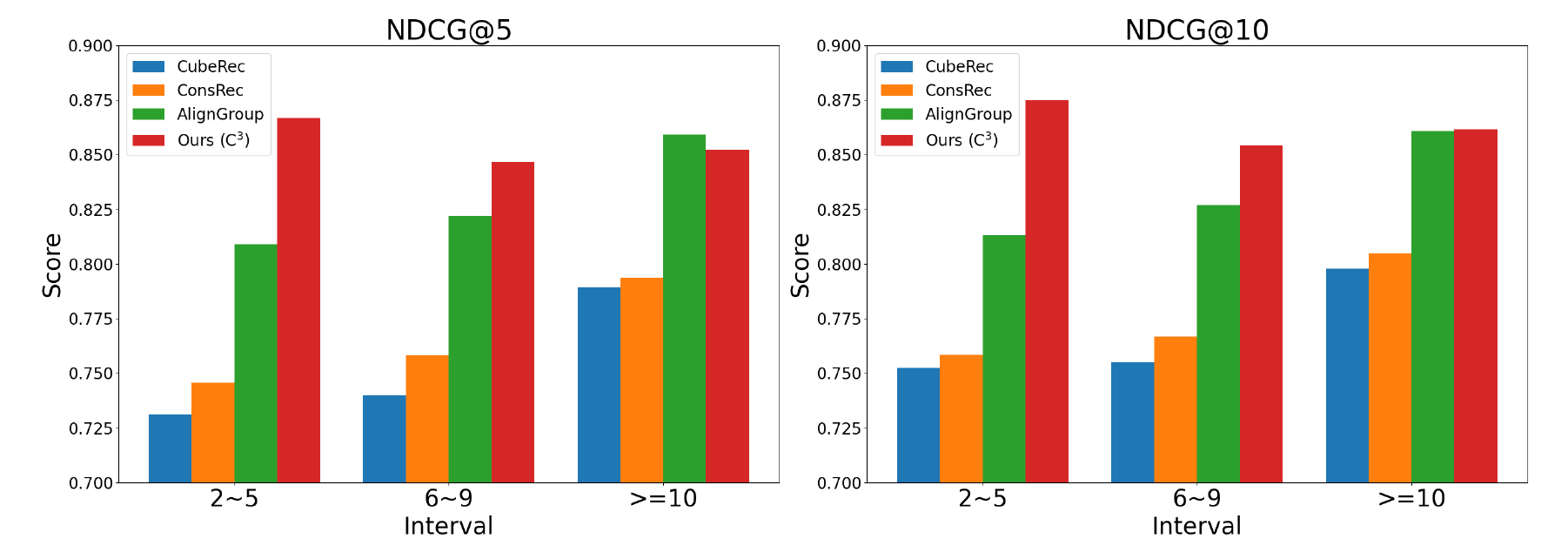}
    \caption{
    NDCG@K comparison across group sizes on the Mafengwo dataset.
}
    \label{fig:rq3}
\end{figure}

In group recommendation, different group sizes affect performance, so robustness across sizes is essential~\cite{chen2022cuberec}. 
We compare C\textsuperscript{3} with four baselines on the Mafengwo, categorizing group sizes following the CubeRec~\cite{chen2022cuberec} definition of small groups (i.e., 2$\sim$5 members). As shown in Fig.~\ref{fig:rq3}, the baselines perform worst on small groups and improve with larger sizes, while C\textsuperscript{3} achieves the best performance on small groups and also improves for groups with 10 or more members, demonstrating robustness. A decline for 6$\sim$9 members groups likely stems from their small proportion in the dataset, where even minor errors cause noticeable drops. 

%% file: texs/6-conclusion.tex
\section{Conclusion}
\label{sec:CONCLUSION}
In this paper, we introduce C\textsuperscript{3}, a group recommendation method that captures group consensus using a Transformer encoder. Our method not only enhances group recommendation performance but also improves user recommendation performance, contributing to a more generalized group recommendation system. Additionally, we design a recommendation loss and a contrastive loss to keep group representations robust, preventing the model from focusing too much on highly interactive members and helping it better reflect diverse preferences. Extensive experiments on four public datasets demonstrate the effectiveness of C\textsuperscript{3}.

%% file: texs/7-ack.tex
\subsubsection*{Acknowledgments.}
This research was supported by the Basic Science Research Program of the National Research Foundation (NRF) funded by the Korean government (MSIT) (No. IITP-2026-RS-2024-00346737); and by three additional grants from the Institute for Information \& Communications Technology Planning \& Evaluation (IITP) funded by the Ministry of Science and ICT (MSIT), Korea, through the Global Scholars Invitation Program (No. IITP-2026-RS-2024-00459638), the Graduate School of Metaverse Convergence at Sungkyun-kwan University (No. IITP-2026-RS-2023-00254129), and the ICT Challenge and Advanced Network of HRD (ICAN) support program (No. IITP-2026-RS-2023-00259497).

\subsubsection*{Disclosure of Interests.}
The authors have no competing interests to declare that are relevant to
the content of this article.